\documentclass{amsart}

\usepackage{amsmath}
\usepackage{amsfonts,amssymb}
\usepackage{color}
\usepackage{graphicx,epstopdf}
\title{Limit order trading with a mean reverting reference price}

\author{Saran Ahuja}
\address{Department of Mathematics, Stanford University, Stanford, CA 94305}
\email{ssunny@stanford.edu}

\author{George Papanicolaou}
\address{Department of Mathematics, Stanford University, Stanford, CA 94305}
\email{papanico@math.stanford.edu}

\author{Weiluo Ren}
\address{Department of Mathematics, Stanford University, Stanford, CA 94305}
\email{weiluo@stanford.edu}

\author{Tzu-Wei Yang}
\address{School of Mathematics, University of Minnesota, Minneapolis, MN 55455}
\email{yangx953@umn.edu}

\begin{document}
\maketitle

\begin{abstract}
Optimal control models for limit order trading often assume that the underlying asset price is
a Brownian motion since they deal with relatively short time scales. The resulting optimal bid and ask limit
order prices tend to track the underlying price as one might expect. This is indeed the case with
the model of Avellaneda and Stoikov (2008), which has been studied extensively. We consider
here this model under the condition when the underlying price is mean reverting. Our main result
is that when time is far from the terminal, the optimal price for bid and ask limit orders is constant,
which means that it does not track the underlying price. Numerical simulations confirm
this behavior. When the underlying price is mean reverting, then for times sufficiently far from terminal, it is more advantageous to focus on the mean price and ignore fluctuations around it. 
Mean reversion suggests that limit orders will be executed with some regularity, and this is why they are optimal. We also explore intermediate time regimes where limit order prices are
influenced by the inventory of outstanding orders. The duration of this intermediate regime depends
on the liquidity of the market as measured by specific parameters in the model.

\end{abstract}


\section{Introduction}

Limit orders play an essential role in today's financial markets. How to optimally submit limit orders has therefore become an important research area. Limit order traders set the price of their orders, and the market determines how fast their orders are executed. Avellaneda and Stoikov proposed a stochastic control model \cite{Avellaneda06high-frequencytrading} for a single limit order trader that optimizes an expected terminal utility of portfolio wealth. In this model, market orders are given by
a Poisson flow with rate $A\exp(-\kappa\delta)$ where $\delta$ is the spread between the limit order price and the observed underlying reference price, while $A$ and $\kappa$ are two positive parameters that control the speed of execution, reflecting in this way the liquidity of the market. The assumption of a Poisson flow is based on two empirical facts presented and discussed in \cite{gabaix2006institutional, PhysRevE.62.R4493, maslov2001price, Potters2003,Weber2005}. One is that in equity markets the distribution of the size of market orders is consistent with a power law, and the other is that the change in the depth of the limit order book caused by one market order is proportional to the logarithm of the size of that order. The Avellaneda-Stoikov model is formulated as a stochastic optimal control problem where the trader balances limit order prices and trading frequency to maximize the expected exponential terminal utility of wealth.

The approach of Avellaneda and Stoikov has been analyzed and extended in \cite{buyLowSellHigh,OU_referencePrice,GueantLiquidation,InventoryRisk,gueant2015general,zhang2009two}. 
In this paper, we use the same optimal control problem, but we are interested in
longer time scales. On a short time scale, the reference price can be modeled by a Brownian motion as seems appropriate in high frequency trading. On a longer time scale corresponding to intermediate trading frequency, we may assume a mean reverting reference price modeled by an Ornstein-Uhlenbeck (OU) process. Reviews of mean reverting behavior in equity markets and associated time scales are presented in \cite{fouque11,Hillebrand2005121}. 

In this paper, we present a numerical study of the long-time limit of the optimal limit order prices in the Avellaneda and Stoikov model with an OU price process. In addition, we study analytically the equilibrium value function of the optimal control problem. Long time behavior of a limit order control problem is studied by Gueant, Lehalle, and Fernandez-Tapia \cite{InventoryRisk}. They use the Avellaneda and Stoikov model but with a Brownian motion price process instead of a mean reverting one. They impose inventory limits which, after some transformations, reduce the problem to a finite-dimensional system of ordinary differential equations. They show that the optimal spreads converge to inventory-dependent limits when time is far away from terminal. Zhang \cite{zhang2009two} and Fodra and Labadie \cite{OU_referencePrice, OU_referencePrice2} also study the Avellaneda and Stoikov model with an OU price process, although they do not consider the long time limit of the trader's optimal strategy. Fodra and Labadie analyze the case where the reference price is away from its long term mean. The trader then anticipates and takes advantage of the tendency of the price to go back to the long term mean. In this paper, we are interested in how the trader would behave if he/she expects that the reference price is likely to oscillate around its long term mean for a relatively long time. We study the case in which the trading period consists of multiple mean reversion cycles of the reference price, while Fodra and Labadie \cite{OU_referencePrice} consider one or just a half of such a cycle.

Our main result is that the optimal limit order prices, instead of the optimal spreads, converge to limits that are independent of all the state variables in the model. This is shown
numerically by two different computational methods. The limit value function is also studied analytically which confirms the accuracy and stability of our numerics. In addition, we observe numerically that the speed at which the optimal limit order prices become insensitive to the reference price is different from that of the inventory levels where the former converges much faster. When the trading period is sufficiently long, there are three stages in the optimal trading strategy:

\begin{enumerate}
\item Far from the terminal time, the trader uses constant limit order prices to generate profit with little concern for risk aversion or leftover inventory.
\item At intermediate times, the trader maintains inventory levels by posting limit orders that depend on inventory levels.
\item Near the terminal time, the trading behavior is mostly determined by the exponential utility function.
\end{enumerate}

We observe that in certain parameter regimes when time is away from terminal by several mean reversion cycles of the reference price, the trader updates limit order prices only according to the change of inventory levels independent of the reference price. These changes become smaller as time moves backwards and are effectively zero when time is far away from the terminal time, in which case the trader posts constant limit order prices. Near the terminal time, the optimal limit order prices are affected by the long-term variance of the reference price and the exponential terminal utility function. We also observe that, with other parameters fixed, the optimal limit order prices converge to their long-term limits faster when the market has more liquidity, which in this model is controlled by parameters in the Poisson flow of orders.

When the trader posts constant limit order prices, then wealth accumulates from the difference between the buy-sell limit order prices instead of from the spread, namely the difference between these prices and the reference price. This strategy is somewhat analogous to a pairs trading strategy,  and when the trading period is long enough, it appears to beat the strategy of tracking the reference price. However, by posting constant limit order prices, the trader gives up the ability to control the trading rate, which is determined entirely by the fluctuations of the reference price. As a result, the variance of the inventory is large which is not desirable towards the end of the trading period due to the terminal exponential utility. Therefore, before getting close to the end of the trading period, the trader needs to keep track of the reference price so as to control the trading flow and avoid a large leftover inventory.

By linearizing the exponential trading intensity, the Avellaneda and Stoikov model with an OU reference price is reduced to a model that can be solved analytically. This is done in Zhang \cite{zhang2009two} and also in Fodra and Labadie \cite{OU_referencePrice}. We compare our numerical solutions with the approximation in Zhang \cite{zhang2009two} and find good agreement  when time is not too far away from terminal.

The structure of this paper is as follows: We first present the model in section 2, then introduce the numerical methods used in section 3. The numerical methods are discussed in more detail in the appendix. In section 4 and 5, we discuss our results for the long-time behavior of the optimal limit order prices and compare them with what is expected analytically. We do not have a full analytical treatment of the long-time behavior of the HJB equation at present. However, in section 6, we carry out an equilibrium analysis on the (time-independent) HJB equation and compare the analytical results obtained with those of our long-time numerical simulations. The result confirms the accuracy and stability of our numerical methods.

\section{Trading model}\label{sec:Trading Model}
\subsection{Settings}
We assume that the reference price $S_{t}$ of the risky asset follows an Ornstein-Uhlenbeck (OU) process
\begin{equation}\label{dynamic for s}
	dS_t = \alpha (\mu-S_t) dt+ \sigma dB_t
\end{equation}
where $\alpha$ is the mean-reverting rate, $\mu$ is the long-term mean, and $\sigma$ is the volatility. Note that we are not considering any feedback effect of traders' behavior on the reference price here.

The portfolio of the limit trader consists of two parts: cash and the risky asset. We denote the cash process by $X_t$ and the inventory process of the risky 
asset by $Q_t$. The process $Q_t$ can be expressed as the difference of ask and bid limit orders fulfilled up to time $t$, denoted by $Q_t^a$ and $Q_t^b$:
\begin{equation}\label{dynamic for q}
	Q_t = Q^b_t - Q^a_t + q_0,
\end{equation}
assuming that the trader only post limit orders and $q_0$ is the initial inventory. The portfolio is self-financing, so
\begin{equation}\label{dynamic for x}
	dX_t = p^a_t dQ_t^a - p^b_t dQ_t^b,
\end{equation}
where $p^a_t$ and $p^b_t$ are the ask and bid limit prices respectively. Gathering \eqref{dynamic for s}, \eqref{dynamic for q} and \eqref{dynamic for x}, the 
dynamics of variables in our model are
\begin{equation}\label{dynamics}
	dQ_t = dQ^b_t - dQ_t^a,\quad
	dX_t = p^a_t dQ_t^a - p^b_t dQ_t^b,\quad
	dS_t = \alpha(\mu-S_t)dt+ \sigma dB_t.
\end{equation}
Note that $ p^{a}_{t}$ and $ p^{b}_{t}$ are the controls of the limit order trader, while  the processes of the fulfilled limit orders $Q^{a}_{t}$ 
and $Q^{b}_{t}$ may be affected by those limit order prices as well as the reference price $S_{t}$.

Combining empirical results from econophysics in \cite{gabaix2006institutional, PhysRevE.62.R4493, maslov2001price, Potters2003,Weber2005}, Avellaneda and Stoikov 
proposed that the process of the fulfilled limit orders follows a doubly stochastic Poisson process with intensity $Ae^{-\kappa \delta_{t}}$, where $\delta_{t}$ is 
the spread of the limit order at time $t$, and $A$ and $\kappa$ are positive constants characterizing statistically the liquidity of the asset. Namely
\begin{equation}\label{Poisson rate}
	Q^a_t\sim \text{Poi}(Ae^{-\kappa\delta^a_t}),\quad Q^b_t\sim \text{Poi}(Ae^{-\kappa\delta^b_t}),
\end{equation}
where $\delta^{a}_{t}=p^{a}_{t}-S_{t}$ and $\delta^{b}_{t} = S_{t} - p^{b}_{t}$ are the spread of ask and bid limit orders posted at time $t$. 

The trader aims to solve the optimal control problem 
\begin{equation}
	\label{objective function}
	\sup_{\delta^{a}, \delta^{b}}\mathbb{E}\left[-e^{-\gamma W_{T}}\right].
\end{equation}
where $W_{t} = X_{t} + Q_{t}S_{t}$ is the process of total wealth. 

The parameters in our model are
\begin{equation}\label{parameters interpretation}
\begin{cases}
\text{$A$: the magnitude of market order flow;}\\
\text{$\kappa$: dictating the shape of order book;}\\
\text{$\gamma$: risk-aversion factor;}\\
\text{$\alpha$: the mean reverting rate of the reference price;}\\
\text{$\sigma$: the volatility of the reference price;}\\
\text{$T$: the length of the trading period.}
\end{cases}
\end{equation}
\newline
\subsection{Dynamic programming}\label{sec:Dynamic Programming}
Consider the value function 
\begin{equation}\label{very first value function}
	u(t, q, x, s) = \sup_{\delta^{a}, \delta^{b}} \mathbb{E}\left[-e^{-\gamma W_{T}} | Q_{t} = q, X_{t} = x, S_{t} = s\right].
\end{equation}
The HJB equation for the optimal control problem specified in \eqref{dynamics} \eqref{Poisson rate} and \eqref{objective function} is 
\begin{equation}
	\label{original HJB}
	\begin{split}
		&u_t + \frac{\sigma^2}{2} u_{ss} + \alpha(\mu - s)u_s
		+ \sup_{\delta^a}\left\{\left[u(t, q - 1, x + s + \delta^a, s) - u(t, q , x, s)\right]Ae^{-\kappa \delta^a}\right\}\\
		&\quad + \sup_{\delta^b}\left\{\left[u(t, q + 1, x - s + \delta^b, s) - u(t, q , x, s) \right]Ae^{-\kappa \delta^b}\right\} = 0
\end{split}
\end{equation}
with the terminal condition $u(T, q, x, s) = -e^{-\gamma(x + q s)}$.

Because of the special form of the terminal utility, namely the CARA\footnote{constant absolute risk aversion} utility, it is known from the studies in Zhang \cite{zhang2009two} and Gueant, Lehalle, and Fernandez-Tapia \cite{InventoryRisk} that the ansatz $u(t, q, x, s) = -e^{-\gamma(x+v(t,q,s))}$ can reduce (\ref{original HJB}) to 
\begin{equation}\label{reduced HJB}
\begin{split}
	&v_t -  \frac{\sigma^2}{2}(\gamma v_s^2 - v_{ss}) + \alpha(\mu-s) v_s\\
	&\quad + \frac{1}{\gamma}\sup_{\delta^a}\left\{[1 - e^{-\gamma(s+\delta^a + v(t,q-1, s) - v(t,q,s))}]Ae^{-\kappa \delta^a}\right\}\\
	&\quad + \frac{1}{\gamma}\sup_{\delta^b}\left\{[1 - e^{-\gamma(-s+\delta^b + v(t,q+1,s) - v(t,q,s))}]Ae^{-\kappa \delta^b}\right\} = 0
\end{split}
\end{equation}
with terminal condition
\begin{equation}\label{terminal condition}
	v(T, q, s) = qs.
\end{equation}
To find the optimal feedback control, we only need to maximize 
\begin{equation}
\begin{split}
	F^a(\delta^a) &= \left[1 - e^{-\gamma(  s+ \delta^a + v(t, q-1, s) - v(t,q,s) )}\right]Ae^{-\kappa \delta^a}\\
	F^b(\delta^b) &= \left[1 - e^{-\gamma( -s+ \delta^b + v(t, q+1, s) - v(t,q,s) )}\right]Ae^{-\kappa \delta^b}
\end{split}
\end{equation}
separately. Both $F^{a}$ and $F^{b}$ have a unique global maximum which yields the optimal feedback spreads
\begin{equation}\label{feedback control spread}
\begin{split}
	&\delta^{a*}(t,q,s) = \frac{1}{\gamma}\log\left(1+\frac{\gamma}{\kappa}\right)  - s - v(t, q-1, s) + v(t, q, s),\\
	&\delta^{b*}(t,q,s) = \frac{1}{\gamma}\log\left(1+\frac{\gamma}{\kappa}\right)  + s - v(t, q+1, s) + v(t, q, s).
\end{split}
\end{equation}
Therefore the problem is reduced to solving the HJB equation 
\begin{equation}\label{reducedHJB_afterPlugIn}
\begin{split}
	&v_t -  \frac{\sigma^2}{2}(\gamma v_s^2 - v_{ss}) + \alpha(\mu-s) v_s
	+ \frac{1}{\gamma}[1 - e^{-\gamma(s+\delta^{a*} + v(t,q-1, s) - v(t,q,s))}]Ae^{-\kappa \delta^{a*}}\\
	&\quad + \frac{1}{\gamma}[1 - e^{-\gamma(-s+\delta^{b*} + v(t,q+1,s) - v(t,q,s))}]Ae^{-\kappa \delta^{b*}} = 0
\end{split}
\end{equation}
with the terminal condition in (\ref{terminal condition}) and the optimal controls in (\ref{feedback control spread}).

We make a change of time  $\tau := T-t$ in  (\ref{reducedHJB_afterPlugIn}), and define $\tilde{v}(\tau, q, s) = v(T-t, q, s)$. 
We abuse the notation by still using $v$ instead of $\tilde{v}$. Plugging the optimal controls to (\ref{feedback control spread}), we have
\begin{equation}\label{time dependent PDE}
\begin{split}
	v_{\tau} &=  \frac{\sigma^2}{2}(v_{ss} - \gamma v_s^2) + \alpha(\mu - s) v_s
	+\frac{A}{\kappa +\gamma} \left(1+\frac{\gamma}{\kappa}\right)^{-\frac{\kappa}{\gamma}}e^{-\kappa(-s - v(\tau, q-1, s) + v(\tau, q, s))}\\
	&\quad + \frac{A}{\kappa +\gamma} \left(1+\frac{\gamma}{\kappa}\right)^{-\frac{\kappa}{\gamma}}e^{-\kappa(s - v(\tau, q+1, s) + v(\tau, q, s))}
\end{split}
\end{equation}
with the initial condition $v(0, q, s) = qs$.

Note that \eqref{time dependent PDE} is highly nonlinear because of the appearance of value function $v$ in the exponent. Moreover, this equation involves both continuous variables, $t$ and $s$, and a discrete variable $q$. There is no available theory on its well-posedness. On the other hand, for the case $\alpha=0$, this equation can be transformed to an ODE system, which, under the assumption of finite inventory limits, is finite-dimensional and can be solved explicitly. See Zhang \cite{zhang2009two} or Gueant, Lehalle and Fernandez-Tapia \cite{InventoryRisk} for detail.

\subsection{Scaling}\label{sec:scaling}

We use two scalings for our model, one on time and another one on price: $\tilde{t} = t/\alpha$ and $\tilde{p}=\gamma p$.
We define new variables in our model
\begin{equation*}
	\tilde{Q}_t = Q_{\tilde{t}/\alpha},\quad \tilde{X}_t = \gamma X_{\tilde{t}/\alpha},\quad \tilde{S}_t = \gamma S_{\tilde{t}/\alpha},\quad
	\tilde{\delta}^a_t = \gamma \delta^a_{\tilde{t}/\alpha},\quad \tilde{\delta}^b_t = \gamma \delta^b_{\tilde{t}/\alpha}.
\end{equation*}
and new parameters accordingly
\begin{equation}\label{new parameters after two scalings}
\tilde{A} = \frac{A}{\alpha},\quad
\tilde{\sigma} = \gamma\frac{\sigma}{\sqrt{\alpha}},\quad
\tilde{\mu} = \gamma\mu,\quad
\tilde{\kappa}  = \frac{\kappa}{\gamma},\quad
\tilde{T} = \alpha T.
\end{equation}
We will use those variables and parameters from now and abuse the notations by dropping all the tildes. With those new variables and parameters, the model is equivalent to that described in section \ref{sec:Trading Model} with $\alpha=\gamma=1$, and the HJB equation \eqref{time dependent PDE} becomes 
\begin{equation}
	\label{time dependent PDE after scaling}
	\begin{split}
		v_{\tau} &=  \frac{\sigma^2}{2}(v_{ss} - v_s^2) + (\mu - s) v_s
		+ \frac{A}{\kappa + 1} \left(1+\frac{1}{\kappa}\right)^{-\kappa} e^{-\kappa(-s - v(\tau, q-1, s) + v(\tau, q, s))}\\
		&\quad + \frac{A}{\kappa + 1} \left(1+\frac{1}{\kappa}\right)^{-\kappa}e^{-\kappa(s - v(\tau, q+1, s) + v(\tau, q, s))}
	\end{split}
\end{equation}
with the initial condition $v(0, q, s) = qs$.

The optimal feedback controls are given by
\begin{equation}\label{optimal feedback control after scaling}
\begin{split}
	&\delta^{a*}(t,q,s) = \log\left(1+\frac{1}{\kappa}\right) - s - v(t, q-1, s) + v(t, q, s)\\
	&\delta^{b*}(t,q,s) = \log\left(1+\frac{1}{\kappa}\right) + s - v(t, q+1, s) + v(t, q, s).
\end{split}
\end{equation}
From now on, we will only consider the scaled model, the HJB equation in \eqref{time dependent PDE after scaling}, and optimal feedback controls in \eqref{optimal feedback control after scaling}. However, when showing our numerical simulation results, we would use the prices before the price-scaling, which are directly observable from the market, instead of the dimensionless ones after the scaling. For the parameters used in the numerical simulations, we may also choose the ones before the price-scaling, since they are easier to reason and are practically easier to calibrate to market data.

We point out that
\begin{enumerate}
\item After scaling, the price-related quantities $S_{t}$, $X_{t}$, $\delta^{a}_{t}$, $\delta^{b}_{t}$, $\mu$ and $\sigma$ are not observable as in \eqref{dynamics} and \eqref{Poisson rate}. Instead, they are dimensionless and measured in the scale of the trader's risk aversion level.
\item The function $v$ and variable $s$ in \eqref{time dependent PDE after scaling} and \eqref{optimal feedback control after scaling} are actually $\gamma v$ and $\gamma s$ in terms of $\gamma$, $v$, and $s$ before the price-scaling. \item The optimal controls in \eqref{optimal feedback control after scaling} are the ones in \eqref{feedback control spread} scaled by $\gamma$.

\end{enumerate}

In the subsequent sections, when discussing how the parameters would affect the model, we will be referring to the new parameters after the scaling instead of those in \eqref{parameters interpretation}. Note that even though we dropped two parameters, namely $\alpha$ and $\gamma$, we have not lost any generality after those two scalings. For a model in \eqref{dynamics}, \eqref{Poisson rate}, and \eqref{objective function} with an arbitrary group of parameters, we can solve a model with scaled parameters constructed in \eqref{new parameters after two scalings}, then convert it to a solution of the original model before scalings.

\section{Numerical methods}\label{sec:Numerical Method}



We briefly discuss two numerical methods that we will use to solve the optimal stochastic control problem described in section  \ref{sec:Trading Model}, particularly equation \eqref{time dependent PDE after scaling}, and produce all the results discussed in the subsequent sections.

The first method is a fully-implicit finite difference scheme. This method has advantages of being relatively simple to implement and  numerically stable. However, it can be slow due to the iteration required at each time steps. 
Secondly, we implement what is called a \textit{split-step scheme} which performs the numerics separately between the linear and nonlinear part of the equation. We briefly describe the second method here and refer to the appendix for more detail on both methods.


We consider the following transformation of the value function $v$ in \eqref{time dependent PDE after scaling}
\begin{equation}\label{relation between v functions}
\tilde{v} = e^{- v}
\end{equation}
which satisfies
\begin{equation}\label{PDE for w}
\begin{cases}
	\tilde{v}_\tau = (\mu - s)\tilde{v}_s + \frac{\sigma^2}{2} \tilde{v}_{ss} 
	- \frac{A}{\kappa + 1}\left(Ae^{-\kappa \delta^{a*}} + Ae^{-\kappa \delta^{b*}} \right)\tilde{v}\\
	\tilde{v}(0, q, s) = e^{- qs}
\end{cases}
\end{equation}
where
\begin{equation}\label{feedback control for w}
\begin{split}
	&\delta^{a*}(\tau,q,s) =  - s + \left[\log\left(1+\frac{1}{\kappa} \right) + \log\tilde{v}(\tau, q-1, s) - \log\tilde{v}(\tau, q, s)\right]\\
	&\delta^{b*}(\tau,q,s) = s + \left[\log\left(1+\frac{1}{\kappa}\right)  + \log\tilde{v}(\tau, q+1, s) - \log\tilde{v}(\tau, q, s)\right]
\end{split}
\end{equation}

We split the PDE in \eqref{PDE for w} to two PDEs:
\begin{equation}\label{mean reversion part}
	\tilde{v}_\tau = (\mu - s) \tilde{v}_s +\frac{\sigma^2}{2} \tilde{v}_{ss}
\end{equation}
\begin{equation}\label{control part}
	\tilde{v}_\tau = -\frac{A}{\kappa + 1}\left(Ae^{-\kappa \delta^{a*}} + Ae^{-\kappa \delta^{b*}} \right)\tilde{v}.
\end{equation}
 
Here equation \eqref{mean reversion part} can be solved via the Feymann-Kac formula, and equation \eqref{control part} can be solved exactly using the method in Zhang \cite{zhang2009two} if we impose finite inventory limits for our problem, in which case the transformation
\begin{equation}
w(t, s, q) = e^{-\kappa s q}\tilde{v}^{-\kappa} 
\end{equation}
reduces \eqref{control part} to a finite-dimensional ODE system that can be solved using a matrix exponential of a tri-diagnoal matrix. Combining those two steps, we have devised a split-step scheme to solve \eqref{PDE for w}. See the appendix for further details.

The feedback optimal limit prices produced by these two methods match very well if we discretize the time space and reference-price space properly.  Compared to the 
finite difference method, the split-step method is much faster since there is no iteration involved. Moreover, the split-step used the Feymann-Kac formula dealing 
with the mean reversion feature in the model, which is fully implicit, stable, and suitable for observing the long time behavior. However, because of 
\eqref{relation between v functions}, the function $\tilde{v}$ may face an underflow/overflow issue when the absolute value of function $v$ is large, which would be 
the case if we allow large $s$ or $q$ in our computation or use fairly large parameters. Therefore, compared to the split-step method, the finite difference method can be 
applied to a wider range of parameters.

\section{Long time behavior}\label{sec:LongTimeBehavior}
 Studying standard Avellaneda-Stoikov model, Gueant, Lehalle, and Fernandez-Tapia \cite{InventoryRisk} observed a long-term stationary behavior of the optimal spreads $\delta^{a*}_{t}$ and $\delta^{b*}_{t}$:
\begin{equation}\label{optimal spread limits}
	\lim_{T-t\rightarrow \infty}\delta^{a*}(t,q) = \delta^{a*}_{\infty}(q),\quad \lim_{T-t\rightarrow \infty}\delta^{b*}(t,q) = \delta^{b*}_{\infty}(q)
\end{equation}

In our model, we observe a long-time behavior of the optimal limit order prices $p_{t}^{a*} = S_{t} + \delta^{a*}_{t}$ and $p_{t}^{b*} = S_{t} - \delta^{b*}_{t}$ instead of that of the optimal spreads $\delta^{a*}_{t}$ and $\delta^{b*}_{t}$. 

Our numerical simulations presented in section \ref{sec:Numerical results} indicate that the optimal feedback limit order prices given by
\begin{equation}\label{feedback control}
\begin{split}
	&p^{a*}(\tau,q,s) =  \log\left(1+\frac{1}{\kappa}\right) - v(\tau, q-1, s) + v(\tau, q, s) \\
	&p^{b*}(\tau,q,s) = -\log\left(1+\frac{1}{\kappa}\right) + v(\tau, q+1, s) - v(\tau, q, s)
\end{split}
\end{equation}
converge to constants
\begin{equation}\label{optimal price limits}
	p^{a*}_{\infty} = \mu + \log\left(1 + \frac{1}{\kappa}\right),\quad p^{b*}_{\infty} = \mu - \log\left(1 + \frac{1}{\kappa}\right)
\end{equation}
when $\tau\rightarrow \infty$. Or equivalently,
\begin{equation}\label{limiting behavior}
\lim_{\tau \to \infty} v(\tau, q, s) - v(\tau, q-1, s) = \mu.
\end{equation}

We have not yet developed an analytical proof of the convergence in \eqref{limiting behavior} as this is work in progress. Note that $v(0, q, s) = qs$, so $v$ can be 
intuitively viewed as the value of the asset held by the trader at time $T-\tau$. The limiting property in \eqref{limiting behavior} suggests that, in the long run, 
the value of each share of asset is just $\mu$, the long term mean of the reference price. Moreover, the convergence suggests that when we are sufficiently far away 
from the terminal time, it is better to post constant limit prices than to track the reference price closely. One heuristic explanation is that the trading period is 
so long compared to the mean reversion time that plenty of rebalancing is guaranteed. Therefore, as long as the trader can gain the premium from rebalancing by using 
the constant limit ask and bid prices, he/she does not need to track the reference price.

In the rest of this section, we discuss three closely related models that can be solved analytically and compare the limit of the optimal limit prices in those models with the ones in \eqref{optimal price limits}.

\subsection{Model with constant reference price}\label{sec:Model with constant reference price}
In our case, the final limit of the optimal prices does not depend on the long-term standard deviation of the reference price. Instead, it uses the spread 
$\log(1 + 1/\kappa)$ relative to the long-term mean of the reference price. It turns out that this spread is closely related to the model with a constant reference 
price. More specifically, in the same scaled model described in section \ref{sec:scaling} but with constant reference price, that is, 
\begin{equation}\label{constant_st}
	S_t \equiv \mu,
\end{equation}
we can solve \eqref{reduced HJB} with \eqref{constant_st} and see that the constants in \eqref{optimal price limits} are the exact optimal limit prices and $\log(1+1/\kappa)$ is the exact optimal spread in this degenerate case.

\subsection{Analysis of small $\kappa$}\label{sec:Analysis of small kappa}
Both Fodra and Labadie \cite{OU_referencePrice} and Zhang \cite{zhang2009two} considered approximations of \eqref{time dependent PDE} with linearization. We briefly state Zhang's results here. 

After a linearization of the exponential terms, Zhang shows analytically that the value function $v$ becomes independent from $s$ exponentially fast
\begin{equation}\label{Kaiyuan's convergence of v}
	v(\tau, s, q) - C_{0} \tau \longrightarrow \theta_{1} + \mu\cdot q - \frac{\sigma^2}{4}q^2,\quad \tau\longrightarrow\infty,
\end{equation}
where $C_0$ and $\theta_1$ are two constants. Accordingly, the optimal prices converge to the following limits:
\begin{equation}\label{limits of optimal LO prices in Zhang's small kappa analysis}
\begin{split}
	p^{a*}_{\infty}(q) &=  \log\left(1 + \frac{1}{\kappa}\right) + \mu - \frac{\sigma^2}{4}(2q - 1)\\
	p^{b*}_{\infty}(q) &= -\log\left(1 + \frac{1}{\kappa}\right) + \mu - \frac{\sigma^2}{4}(2q + 1),
\end{split}
\end{equation} 
where the slope with respect to $q$ relies only on the scaled $\sigma$. In order for the linearization to work well, the terms $\kappa\delta^{a*}$ and $\kappa\delta^{b*}$ are required to be small. According to \eqref{limits of optimal LO prices in Zhang's small kappa analysis}, the terms $\delta^{a*}$ and $\delta^{b*}$ will be linear with respect to $q$ when time is far away from terminal, so at least in this time regime, to keep $\kappa\delta^{a*}$ and $\kappa\delta^{b*}$ small, this result is only valid for sufficiently small $q$.

\begin{figure}
\centering
\includegraphics[width=0.99\linewidth]{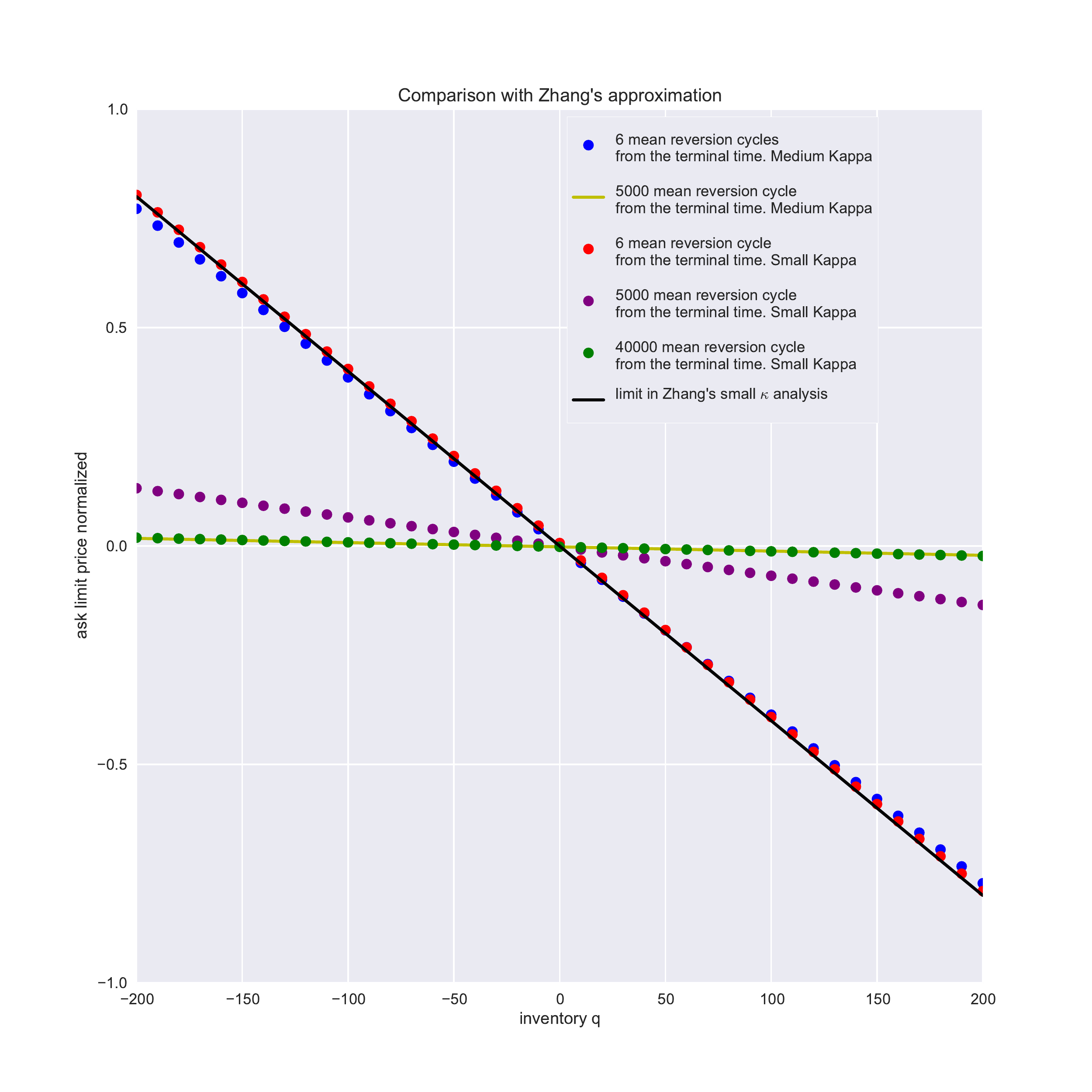}
\caption{We plot optimal ask limit order prices at different times from the model with small $\kappa$ ($\kappa = 1$) and the model with medium $\kappa$ ($\kappa=6$). The black line is the ask price in Zhang's small $\kappa$ analysis for comparison. To make the prices from 
different models comparable, we subtract the optimal feedback ask price at $q=0$ from each feedback ask price function. Note that after this normalization, the limit 
of Zhang's small $\kappa$ approximation from two models coincides with each other.\newline
When time is sufficiently far away from the terminal time, the optimal ask limit order prices from both models become significantly different from the limit in Zhang's small $\kappa$ analysis and converge a constant. Moreover, the limit order price in the model with small $\kappa$ tends to a constant more slowly than the one in the model with medium $\kappa$.}
\label{fig:KaiyuanApproxCmp}
\end{figure}

We compare the optimal feedback ask limit prices computed by our numerical methods to those in the limit of Zhang's approximation\footnote{The shared parameters are $A=10$ and $\sigma=0.02$. We considered one model with medium $\kappa$ ($\kappa = 6$) and another one with small $\kappa$ ($\kappa = 1$).} in Figure \ref{fig:KaiyuanApproxCmp}. We plot the feedback ask limit prices as functions of inventory $q$ as they have already become insensitive to the reference price $s$. Translation is applied on those feedback optimal prices to make them comparable. We refer to Figure \ref{fig:KaiyuanApproxCmp} for more detail.


\subsection{Model with linear utility}
When time is far away from the terminal time, the trader has little pressure from risk aversion rooted in the terminal exponential utility, so we expect the trading pattern in such a scenario to be similar to the one in the model with linear utility
\begin{equation}
	\mathbb{E}[X_T+Q_T S_T],
\end{equation}
which can be view as a degenerate case of the model with the exponential utility when $\gamma\to 0$.

In \cite{OU_referencePrice}, Fodra and Labadie have considered this case and have obtained the analytical solution for the optimal prices. In this case, the optimal feedback limit order prices would converge exponentially fast to 
\begin{equation}\label{limit in linear utility model}
\hat{p}^{a*}_{\infty} = \mu + \frac{1}{\kappa},\quad \hat{p}^{b*}_{\infty} = \mu - \frac{1}{\kappa}
\end{equation}

Recall that the limits of the dimensionless optimal feedback prices in our model with exponential utility are
\begin{equation}
p^{a*}_{\infty} = \mu + \log\left(1 + \frac{1}{\kappa}\right),\quad p^{b*}_{\infty} = \mu - \log\left(1 + \frac{1}{\kappa}\right).
\end{equation}
If we do not scale the price-related quantities by $\gamma$ in section \ref{sec:Trading Model}, then the limit of the optimal prices, which are dimensional in this case, are
\begin{equation}\label{limit in exponential utility model}
p^{a*}_{\infty} = \mu + \frac{1}{\gamma}\log\left(1 + \frac{\gamma}{\kappa}\right),\quad p^{b*}_{\infty} = \mu - \frac{1}{\gamma}\log\left(1 + \frac{\gamma}{\kappa}\right)
\end{equation}
Note that $1/\kappa$ in \eqref {limit in linear utility model} is just the limit of $\frac{1}{\gamma}\log(1 + \gamma/\kappa)$ in 
\eqref{limit in exponential utility model} when the risk aversion parameter $\gamma\to 0$, so the limits derived from those two models are consistent.

In the linear utility case, the constant strategy is almost optimal when $T-t$ becomes greater than several mean reversion cycles of the reference price. Since the linear utility is a degenerate case of the exponential utility, it is not surprising that in the exponential utility case, the strategy with constant limit prices also becomes optimal when $T-t$ is large.

\section{Numerical results}\label{sec:Numerical results}
We apply the numerical methods described in section \ref{sec:Numerical Method} to solve the HJB equation \eqref{time dependent PDE after scaling} for the value 
function and optimal controls in our model.

\subsection{Evolution of optimal feedback limit order prices}
\begin{figure}
\centering
\includegraphics[width=0.75\linewidth]{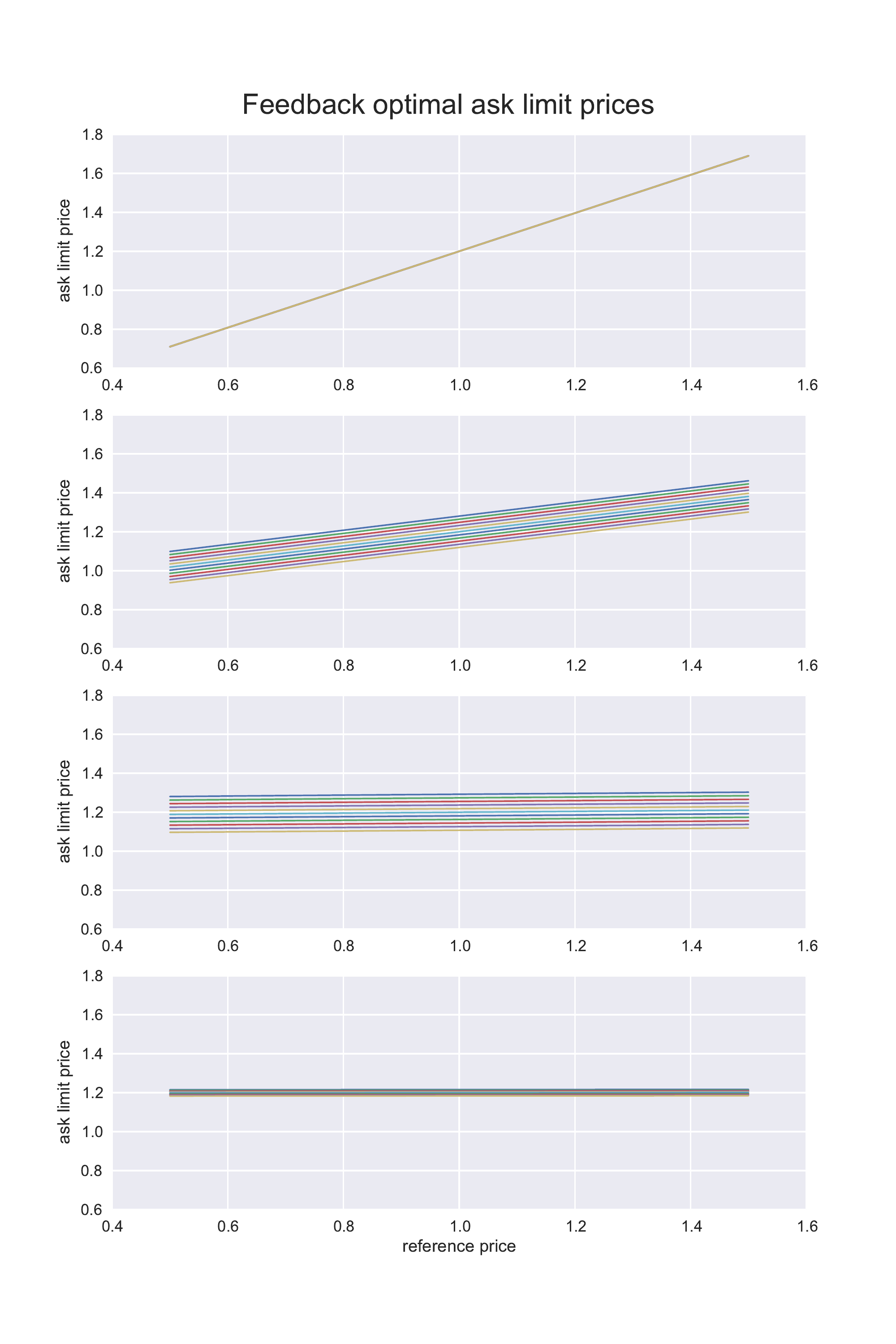}
\caption{Feedback optimal ask limit order prices, from top to bottom, corresponding to 0, 1, 4 and 800 mean reversion cycles from the terminal time.
The prices are the ones before the price-scaling described in section \ref{sec:scaling} instead of the dimensionless ones after the scaling. Each line, as a function 
of reference price, corresponds to a value of inventory. The optimal ask prices have already become independent from the reference price at 4 mean reversion cycles 
from the terminal time (the 3rd plot from top), while it took 800 mean reversion cycles (backwards in time) to become independent from the inventory as well (the bottom plot). Here the 3rd plot from top corresponds to the intermediate regime and the bottom plot corresponds to the far-away-from-terminal regime.}
\label{fig:optimalAskPriceSeq}
\end{figure}

We are interested in how the optimal feedback controls in our model, namely the optimal limit order prices, evolve as a function of the inventory and reference 
price. As stated in section \ref{sec:LongTimeBehavior}, we observe that the optimal prices converge to constants in \eqref{optimal price limits} when time is away from terminal. In addition, as shown in Figure \ref{fig:optimalAskPriceSeq}, we observe that the optimal limit order prices become insensitive to the reference price much faster than to the inventory, which leads to an ``intermediate'' regime where 
the optimal limit order prices only respond to the change of inventory.

For a model with unscaled parameters $A= 10$, $\sigma = 0.05$, $\gamma=0.005$, $\kappa=5.0$, $\mu=1$ and $\alpha=1$, Figure \ref{fig:optimalAskPriceSeq} shows the optimal feedback ask prices at following time:

\begin{enumerate}
\item the terminal time;
\item near-terminal regime: 1 mean reversion cycle of the reference price from the terminal time; 
\item intermediate regime: 4 mean reversion cycles from the terminal time;
\item limit regime: 800 mean reversion cycles from the terminal time.
\end{enumerate}

We can see that the near-terminal regime is short and very quickly, backwards in time, trading gets into the intermediate regime where the optimal prices become insensitive to the reference price. In the intermediate regime, the trader updates his limit order only according to the inventory. Doing so, he could keep the variance of inventory low which reflects his risk aversion rooted in the terminal utility.

In Figure \ref{fig:optimalAskPriceSeq}, it takes 800 mean reversion cycles to observe the insensitivity of the optimal prices to the inventory, as shown in the bottom plot. Namely for a large portion of a trading period, the trader would post limit orders with prices only affected by the change of his own inventory ignoring the fluctuation of the reference. 

In some parameter ranges, the intermediate regime can be very long and so we observe insensitivity of optimal limit prices to the reference price but do observe dependence on inventory. That is, in the very beginning of the trading period the intermediate regime is already valid, in which case we would not observe the limit regime at all. To illustrate this, we choose two sets of parameters and show the corresponding simulation results in the next section where within the trading period we can only observe the intermediate and near-terminal regimes but not the limit regime.

Note that even though in this paper we do not calibrate our parameters to real data, there is literature on how to do this. For the liquidity parameters $A$ and $\kappa$, their calibration is studied in Chapter 4 of \cite{fernandez2015modeling}. A calibration framework is presented, which can be extended to the model with a mean reverting reference price. For the parameters $\alpha$ and $\sigma$ characterizing the mean reverting reference price, which we assume is observed, we can calibrate them by a maximal likelihood estimation (MLE) studied in \cite{mullowneymaximum, valdivieso2009maximum}. 


\subsection{Simulation results}\label{sec:Simulation results}

\begin{figure}
\centering
\includegraphics[width=0.90\linewidth]{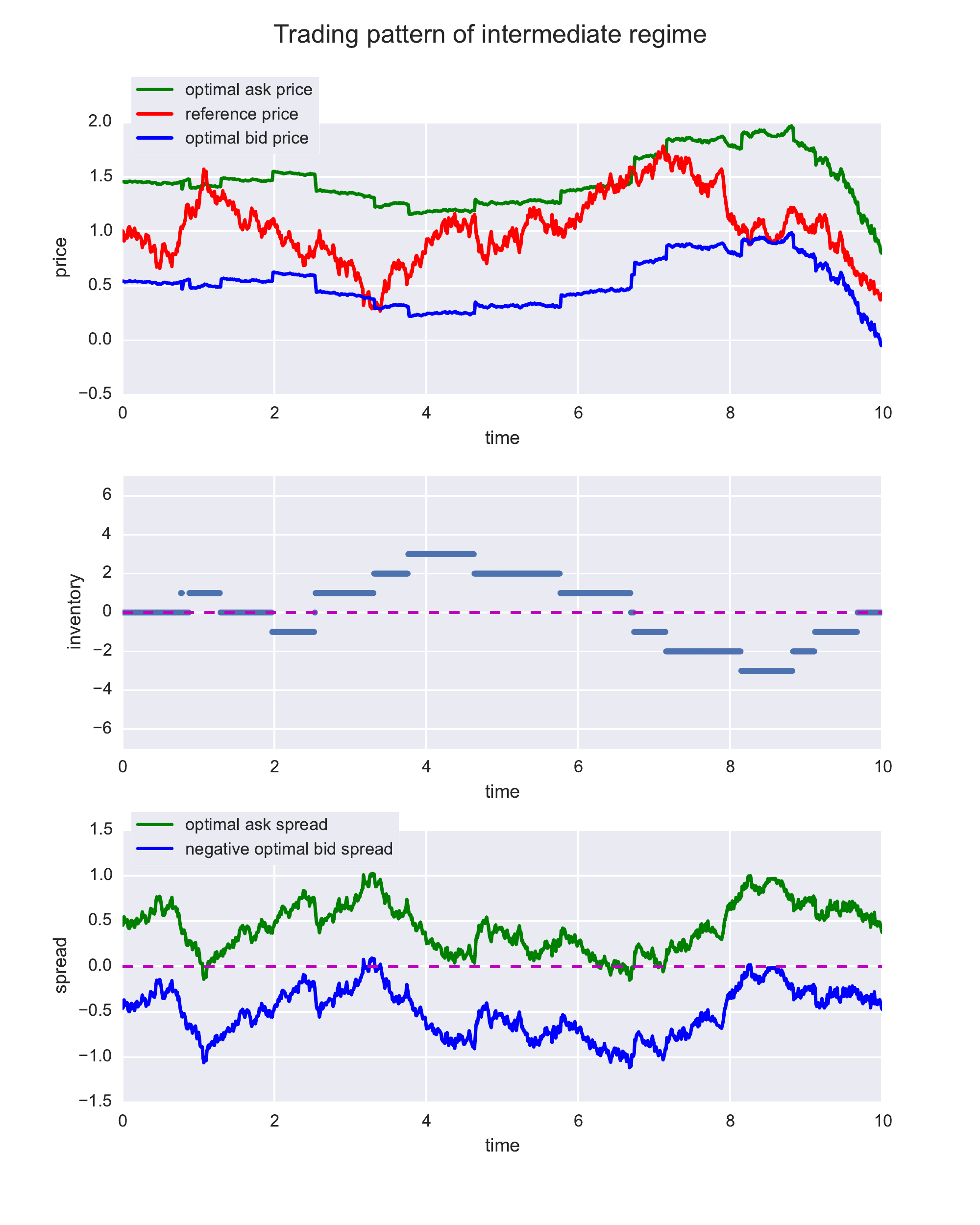}
\caption{Simulation results for limit order prices, inventory, and spreads for 10 mean reversion cycles of the underlying reference price. The pattern clearly shows that near the terminal time, the trader tracks the reference price closely whereas in the intermediate regime, the optimal limit prices effectively only respond to the change of inventory.}
\label{fig:simulationCmpMediumA}
\end{figure}

\begin{figure}
\centering
\includegraphics[width=0.90\linewidth]{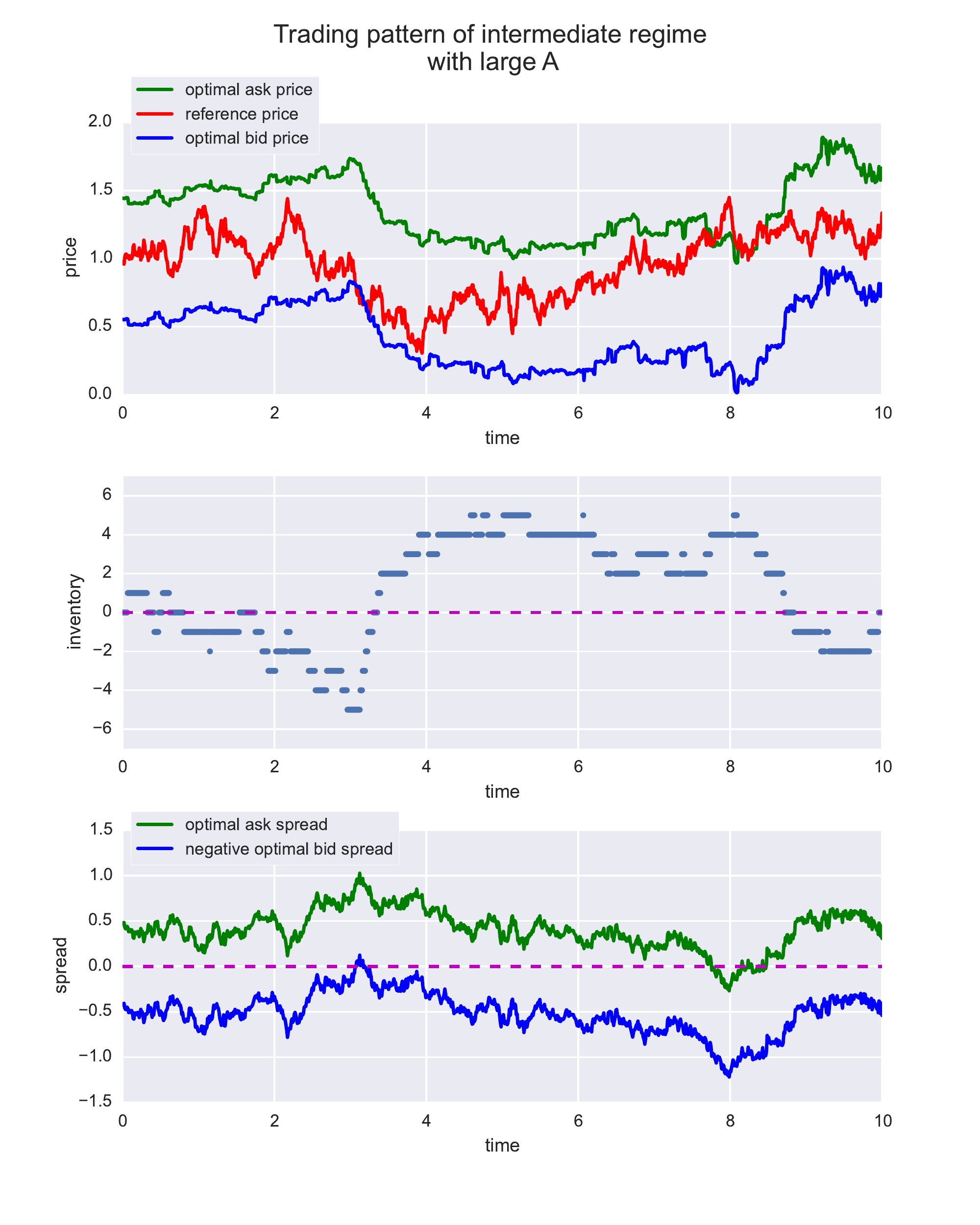}
\caption{Similar plots to Figure \ref{fig:simulationCmpMediumA}, but with larger parameter $A$ representing greater volume of incoming market orders. When there is a trend in the reference price, for instance, between time 2 and 4, there will also be a trend in optimal prices in the same direction but with a lag. The trend in optimal prices is a result of the trend in the inventory formed during a trend of the reference price when the volume of incoming market orders is large.}
\label{fig:simulationCmpLargeA}
\end{figure}

We show some simulation results of our trading models in Figure \ref{fig:simulationCmpMediumA} and Figure \ref{fig:simulationCmpLargeA}. The unscaled parameters used in Figure \ref{fig:simulationCmpMediumA}  are $A=2$, $\sigma=0.4$, $\gamma=2$, $\kappa=1.5$, $\mu=1$ and $\alpha=1$; the ones used in Figure \ref{fig:simulationCmpLargeA} are the same except that $A=6$. We choose those parameters so that the intermediate regime can be observed clearly. For instance, we choose a large $\gamma$ so that a jump of the optimal prices due to a change in inventory is evident. In those two figures, the trading period is not long enough to observe the limit regime.

Figure \ref{fig:simulationCmpMediumA} shows a simulation result for 10 mean reversion cycles of reference price. Between time 0 and 8, the 
trading is in the intermediate regime in the sense that the optimal limit order prices will remain almost constant when no limit order is taken and will jump when 
the inventory changes. 

In Figure \ref{fig:simulationCmpLargeA}, the model has the same parameters except that the market-order-volume parameter $A$ is greater. In the top plot of Figure \ref{fig:simulationCmpLargeA}, while it seems that the optimal prices are tracking the reference 
price, however, a closer look shows that the pattern in this plot is essentially the same as the pattern in the top plot of Figure 
\ref{fig:simulationCmpMediumA}. That is, the limit order prices effectively respond only to the change of inventory and ignore the 
fluctuation of the reference price, which suggests that we are in the ``intermediate regime.'' For instance, between time 2 and 3 in the top plot of Figure \ref{fig:simulationCmpLargeA}, there is a significant drop of the 
reference price, but the limit prices does not drop accordingly. They begin to decrease only after the inventory increases. In this case, parameter $A$ is sufficiently large 
that enough limit orders will be taken in one trend of price, which builds up a trend in the inventory and in turn creates a trend in 
the optimal limit order prices. This explains why on first sight, the limit order prices follow the same trend as the reference price, and why there is a lag between 
the trend of the reference price and that of the limit order prices.

When the model moves from the near-terminal regime to the intermediate regime, the sensitivity of the optimal prices to the inventory is mainly affected by the 
scaled $\sigma$. We observed that the greater the scaled $\sigma$ is, the greater the jump size of the optimal prices is when the inventory changes by one unit. The magnitude of a jump decays to 0 as time goes backwards, with the decay rate affected by $A$ and $\kappa$; for greater values of $A$ and 
$\kappa$, the jump size decays faster. Note that, larger values of $A$ and $\kappa$ means a larger market order flow and a shallower order book respectively. These properties signify higher liquidity in the market.  So one insight we can gain from this model is that, for a limit order trader trading a liquid asset with 
mean-reverting price, his optimal limit prices converge faster backwards in time than they do in the case where he trades a less liquid asset, and therefore his optimal limit prices are less sensitive to the change of inventory. 

%
%
Recall that here the parameters are the ones after scalings described in section \ref{sec:scaling}, so $A$, $\kappa$, and $\sigma^{2}$ are in fact $\frac{A}{\alpha}$, $\frac{\kappa}{\gamma}$, and $\frac{\gamma^{2}\sigma^{2}}{\alpha}$ in terms of the parameters before scalings. In contrast, the prices in the figures shown in this section are those before the price-scaling described in section \ref{sec:scaling}. That is, they are the observable prices instead of the dimensionless ones.

\section{Equilibrium analysis}
To check whether our numerical solution of the system in \eqref{time dependent PDE after scaling} is still valid even when time is far away from terminal, we analytically consider the equilibrium of that system and compare the result with our numerical solution of the time-dependent system.

As described in section \ref{sec:LongTimeBehavior}, our conjecture is that, for a solution $v$ of the PDE \eqref{time dependent PDE after scaling} and any $q$ and $s$,
\begin{equation}
	v(\tau, q, s) - v(\tau, q-1, s) \to \mu,\quad \text{as $\tau\to\infty$}.
\end{equation}
Thus, for the equilibrium, we expect 
\begin{equation}\label{convergence of v}
	v(\tau, q, s) - C\tau \to \theta_0 + \theta(s) + \mu q, \quad \text{as $\tau\to\infty$}
\end{equation}
where $\theta_0$ is a constant and $\theta(s)$ satisfies the equilibrium HJB Equation
\begin{equation}\label{ergodic ODE}
\begin{cases}
	0 = C + \frac{\sigma^2}{2}(\theta_s^2 - \theta_{ss}) - (\mu - s)\theta_s - 	M[e^{-\kappa(-s + \mu)}  +  e^{-\kappa(s - \mu)}]\\
	\theta(\mu) = 0
\end{cases}
\end{equation}
with $M = \frac{A}{\kappa +1}(1+\frac{1}{\kappa})^{-\kappa}>0$. Therefore, when $\tau$ is large, we expect the ``$s$-dependent'' part of $v$, defined as 
$v(\tau, q, s) - v(\tau, q, \mu)$, to be close to $\theta(s)$, the solution to \eqref{ergodic ODE}. We will transform \eqref{ergodic ODE} to a Schr\"odinger 
eigenvalue problem, then solve it and compare the result to the numerical solution of $v$  when time is away from terminal.

Moreover, recall that Zhang \cite{zhang2009two} has obtained a closed form solution of the linearized model with small $\kappa$. We can compare the limit of a value function in our model in \eqref{convergence of v} and the one in Zhang's small $\kappa$ analysis in \eqref{Kaiyuan's convergence of v}. In both equations, when $\tau=T-t$ is large, the derivative of the value function with respect to $\tau$ is a constant: $C$ in \eqref{convergence of v} and $C_0$ in \eqref{Kaiyuan's convergence of v}. It is shown in  \cite{ren2016}  that 
$$\lim_{\kappa\to 0}C = \lim_{\kappa\to 0} C_0.$$
That is, the constants in the two convergence results are consistent when $\kappa$ is small. We refer to \cite{ren2016} for more detail.


Note that the limit in \eqref{convergence of v} is indeed a solution of the HJB equation in \eqref{time dependent PDE after scaling}, but it does not satisfy the initial condition. Now we analyze equation \eqref{ergodic ODE} to gain some insight into constant $C$ and solution $\theta$.

\subsection{Schr\"odinger equation}
First we could assume $\mu$ = 0, or equivalently we can make a change of variable $s - \mu \rightarrow s$. Then we define

\begin{equation}
m = e^{-\frac{1}{2\sigma^2}s^2 - \theta}
\end{equation}
which satisfies
\begin{equation}\label{Schrodinger equation}
 -m'' + \left[\frac{s^2}{\sigma^4} + \frac{2M}{\sigma^2}(e^{\kappa s} + e^{-\kappa s})\right] \cdot m = \hat{C}\cdot m
\end{equation}
where $\hat{C} = C\frac{2}{\sigma^2} + \frac{1}{\sigma^2}$. Here we have a Schr\"odinger operator
\begin{equation}\label{Schrodinger Operator}
	\hat{L}[m] =  -m'' + \left[\frac{s^2}{\sigma^4} + \frac{2 M}{\sigma^2}(e^{\kappa s} + e^{-\kappa s})\right] \cdot m
\end{equation}
with an unbounded positive potential. Therefore, it has a lower-bounded discrete spectrum (see Theorem 7.3  in \cite{pankov2001introduction} for instance). We are looking for the smallest eigenvalue whose eigenfunction vanishes at infinity and does not change its sign on the real line. See \cite{ren2016} for a heuristic discussion on how the smallest eigenvalue of the Schr\"odinger operator in equation \eqref{Schrodinger Operator} appears in the limit of 
$v$ in \eqref{time dependent PDE after scaling}.

\subsection{Numerical results on the equilibrium equation}\label{sec:Numerical results on the equilibrium Equation}
We would like to solve the equation \eqref{Schrodinger equation} to find the constant $\hat{C}$ that yields a solution which vanishes at infinity and does not 
change its sign. Recall that $\hat{C} = C\frac{2}{\sigma^2} + \frac{1}{\sigma^2}$, where the constant $C$ can be approximated by $v_\tau$ with large $\tau$ as shown 
in equation \eqref{convergence of v}. We search in the neighborhood of $\frac{2}{\sigma^2}v_\tau + \frac{1}{\sigma^2}$ to find the desired constant $\hat{C}$. After we have the 
constant $\hat{C}$, we compute the solution of the following system, which is derived from \eqref{Schrodinger equation},
\begin{equation}\label{ode system for m}
\begin{cases}
	m' = n\\
	n' = \left[\frac{s^2}{\sigma^4} + \frac{2M}{\sigma^2}(e^{\kappa s} + e^{-\kappa s}) -  \hat{C}\right] \cdot m.
\end{cases}
\end{equation}

We need to specify the initial conditions $m(0)$ and $n(0)$. Since equation \eqref{Schrodinger equation} is homogeneous, $m(0)$ could be arbitrary, so we set 
$m(0)=1$ for simplicity. Additionally, we set $n(0) = 0$ due to the symmetry of the equation. 

Consider two models with shared parameters
\begin{equation}
A = 0.9, \quad\sigma = 0.3,\quad \gamma = 0.01,\quad\mu=1.0
\end{equation}
and different values of $\kappa$:
\begin{equation}
\kappa = \begin{cases}
0.3\quad\text{in the first model},\\
0.01\quad\text{in the second model}
\end{cases}
\end{equation}
where the parameters are the ones before the price-scaling.

\begin{figure}
\centering
\includegraphics[width=0.99\linewidth]{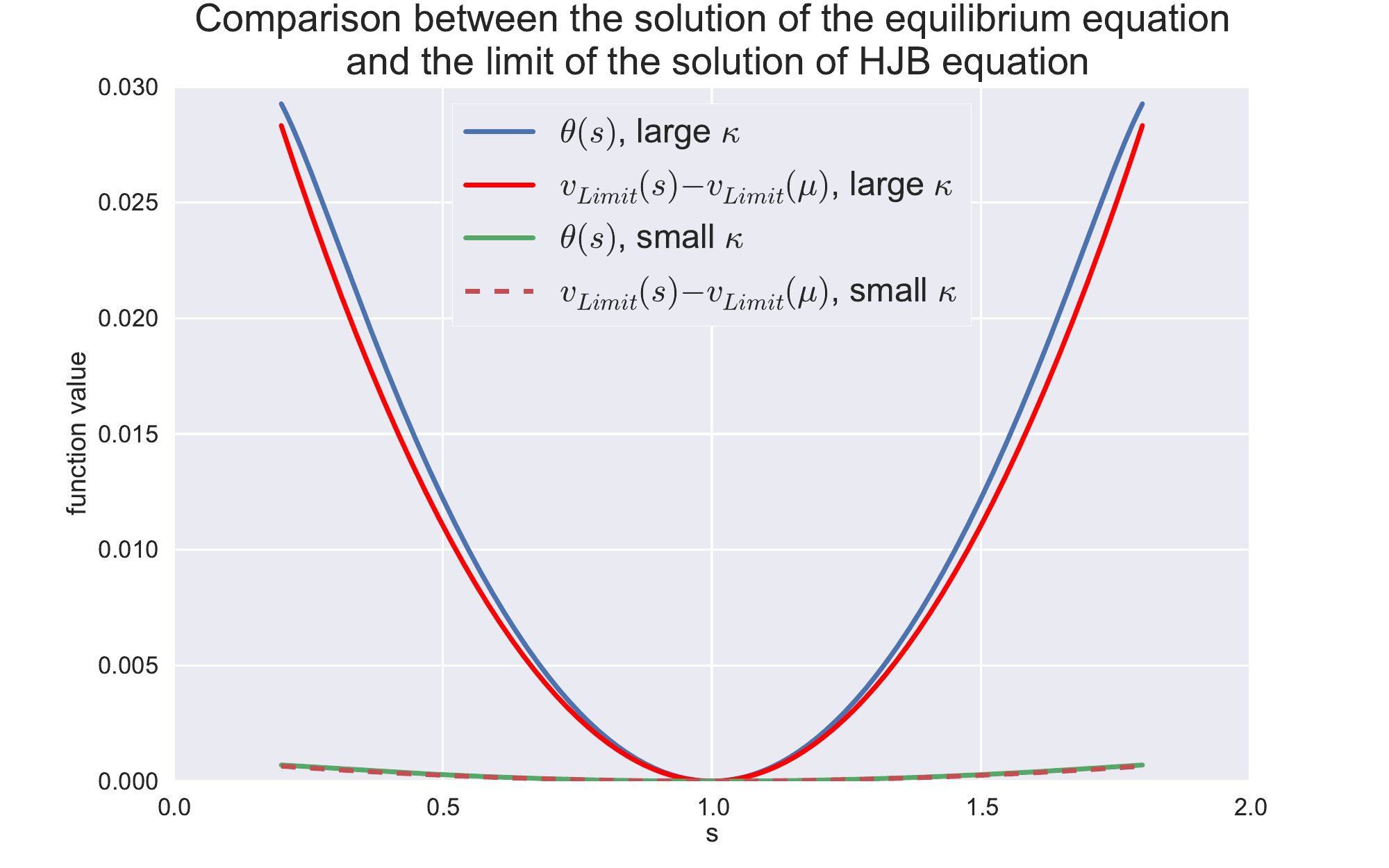}
\caption{We compare $\theta\left(s\right)$ in the Schr\"odinger equation \eqref{ergodic ODE}  and $v_{Limit}\left(s\right) - v_{Limit}\left(\mu\right)$, namely the 
``$s$-dependent'' part of the function $v$ from the numerical result of the time dependent system in equation \eqref{time dependent PDE after scaling} when time is 
far away from terminal. Here $\theta\left(s\right)$ and $v_{Limit}\left(s\right) - v_{Limit}\left(\mu\right)$ match very well for both large $\kappa$ 
($\kappa=0.3$) and the small $\kappa$ ($\kappa=0.01$). When $\kappa$ is small, both $\theta\left(s\right)$ and $v_{Limit}\left(s\right) - v_{Limit}\left(\mu\right)$ 
are flat, which is consistent with the result in Zhang's small $\kappa$ analysis.}
\label{fig:EigenfunctionCmp}
\end{figure}

We numerically solve \eqref{ode system for m} for $m\left(s\right)$, then in turn compute $\theta(s)$ with $\theta(\mu)=0$ in \eqref{ergodic ODE}. We compare it in Figure \ref{fig:EigenfunctionCmp} to the ``$s$-dependent'' part of $v$ which is defined as
\begin{equation}
	v_{Limit}(s) - v_{Limit}(\mu)
\end{equation}
where $v_{Limit}(s)\triangleq v(\tau, q=0, s)$ for large $\tau$. We can see that for each model, $\theta(s)$ is very close to the 
``$s$-dependent'' part of $v$, and that for the model with smaller $\kappa$, $\theta(s)$ is flatter. Note that the result  in section \ref{sec:Analysis 
of small kappa}, which states that the limits of $v$ calculated via a small $\kappa$ expansion does not depend on $s$, can be viewed as the limit when $\kappa$ goes to 0.

\section{Conclusion}\label{sec:Conclusion}	

In this paper, we consider the limit order book model of Avellaneda and Stoikov \cite{Avellaneda06high-frequencytrading} with a mean reverting underlying price. Our 
main result is that when time is far from terminal, it is optimal to post constant limit order prices instead of tracking the underlying price.
We use two different numerical methods to solve the HJB equation, and both of them confirm the long-time behavior. This result implies that 
when the underlying price is mean reverting then, when time is far from terminal, it is optimal to focus on the mean price and ignore the fluctuations around it. 
This observation, admittedly from a stylized model, confirms what limit order traders might expect.

The numerical results also show that between the time regime where constant limit order prices are optimal and the one close to the terminal time, there is an intermediate time period where limit order prices are influenced by the inventory of outstanding orders. The duration of this intermediate period depends on the parameters $A$ and $\kappa$ that quantify the liquidity of the market.

We also study the equilibrium of the optimal control problem. The equilibrium of the HJB equation can be transformed to a Schr\"odinger equation, as an eigenvalue problem. 
The solution agrees with the long-time limit of our numerical result of the time-dependent model, which confirms the validity and accuracy of our numerical methods, even for long time. When the liquidity parameter $\kappa$ is small, the numerical solutions also match the analysis in Zhang \cite{zhang2009two}.

Even though the numerical calculations strongly suggest convergence of the optimal limit order prices, the proof remains open and needs further study. 

\appendix
\appendix

\section{Numerical methods}\label{sec: Numerical methods}
\subsection{Boundary condition}
As in  \cite{InventoryRisk, zhang2009two}, we assume that the total amount of the asset available is $Q$, which means at the boundary point with $q=Q$, buying is forbidden, and at the boundary point with $q=-Q$, selling is forbidden. Those are the boundary conditions at the artificial boundaries $q=\pm Q$ for the inventory space.

It appears that the numerical boundary condition may affect the solution of the optimal control problem described in section \ref{sec:Trading Model} even at the 
points away from the boundary. Here we discuss the model with constant reference price $\mu$ as an example. In section \ref{sec:Model with constant reference price}, 
we have shown that for the model with constant reference price $S_{t} \equiv \mu$, the optimal prices are 
\begin{equation}\label{exact optimal price in model with constant price}
p^{a*} = \mu + \log\left(1+\frac{1}{\kappa}\right),\quad p^{b*} = \mu - \log\left(1+\frac{1}{\kappa}\right).
\end{equation}
However, if we add boundary conditions at the boundaries $q=\pm Q$, then the asymptotic optimal limit bid and ask prices become
\begin{equation}\label{limit with boundary effect}
\begin{split}
	&p^b_\infty(q) = \mu - \log\left(1+\frac{1}{\kappa}\right) 
	+ \frac{1}{\kappa}\left[\log\sin\left(\frac{(q+Q+2)\pi}{2Q+2}\right) - \log\sin\left(\frac{(q+Q+1)\pi}{2Q+2}\right)\right]\\
	&p^a_\infty(q) = \mu + \log\left(1+\frac{1}{\kappa}\right) 
	+ \frac{1}{\kappa}\left[\log\sin\left(\frac{(q+Q+1)\pi}{2Q+2}\right) - \log\sin\left(\frac{(q+Q)\pi}{2Q+2}\right)\right].
\end{split}
\end{equation}
These are different from the exact solution of the problem without boundary conditions in \eqref{exact optimal price in model with constant price}. Note that for 
$|q|\ll Q$, such a difference is indeed negligible.

This stylized example shows how the numerical boundary conditions affect the solution. In practice, we would set $Q$ fairly large and only check the value function or the optimal prices for $|q|<\frac{Q}{4}$ for instance. Note that in \eqref{limit with boundary effect}, with $q$ fixed and $Q\rightarrow \infty$, $p^{a}$ and $p^{b}$ converge to the expression in \eqref{exact optimal price in model with constant price}, so in this degenerate model,
\begin{equation}
	\lim_{t\rightarrow \infty}\lim_{Q\rightarrow \infty} p^b (t, q) = \lim_{Q\rightarrow \infty}\lim_{t\rightarrow \infty} p^b (t, q),\quad
	\lim_{t\rightarrow \infty}\lim_{Q\rightarrow \infty} p^a (t, q) = \lim_{Q\rightarrow \infty}\lim_{t\rightarrow \infty} p^a (t, q).
\end{equation}
We expect that the same result still holds for non-degenerate models. Thus, in practice, we compute $\lim_{t\rightarrow\infty} p^b(t, q)$ and 
$\lim_{t\rightarrow\infty} p^a(t, q)$ for fixed $q$ and large $Q$ to approximate $\lim_{t\rightarrow\infty} \lim_{Q\rightarrow\infty} p^b(t, q)$ and 
$\lim_{t\rightarrow\infty} \lim_{Q\rightarrow\infty} p^a(t, q)$.

\subsection{Finite difference method}

Due to the non-linearity, we implement an implicit finite difference method to solve  \eqref{time dependent PDE after scaling}. We discretize the time space $[0,T]$ and the reference-price space $[\mu - S, \mu + S]$ using step-size $\Delta \tau$ and $\Delta s$,  and consider
\begin{equation*}
	v^n_{q, j} \approx v(n\Delta \tau, q, \mu - S + j\Delta  s)
\end{equation*}
where $n$ is the index of the grid point in the time space, $q$ is the number of asset held, and $j$ is the index of the grid point in the reference-price space. 

At each step, we assume that $v^n_{q, j}$ is known and we compute $v^{n+1}_{q, j}$. The terms in equation \eqref{time dependent PDE after scaling} are 
replaced by the following terms
\begin{equation}
\begin{split}
	&v_\tau \approx \frac{v^{n + 1}_{q, j} - v^{n}_{q, j}}{\Delta \tau},\quad 
	v_s^2 \approx \left(\frac{v^{n+1}_{q, j + 1} - v^{n+1}_{q, j - 1}}{2\Delta s}\right)^2,\quad
	v_{ss} \approx \frac{v^{n+1}_{q, j + 1} - 2v^{n+1}_{q, j} + v^{n+1}_{q, j-1} }{\Delta s^2}\\
	&(\mu - s) v_s \approx (\mu - s)\left[\frac{ v^{n+1}_{q, j + 1} - v^{n+1}_{q, j}}{\Delta s} 1_{\mu > s} 
	+ \frac{ v^{n+1}_{q, j} - v^{n+1}_{q, j-1}}{\Delta s} 1_{\mu < s} \right].
\end{split}
\end{equation}
The discretized PDE can be written as
\begin{equation}\label{finite difference discretized PDE}
\begin{split}
	&v^{n + 1}_{q, j} \left[1 + \frac{\Delta \tau}{\Delta s^2} \sigma^2 
	- (\mu-s)\frac{\Delta \tau}{\Delta s}\left(1_{\mu < s} - 1_{\mu > s}\right)\right]
	+ v^{n + 1}_{q, j+1} \left(-\frac{\Delta \tau}{2\Delta s^2} \sigma^2 - (\mu-s)\frac{\Delta \tau}{\Delta s}1_{\mu > s}\right)\\
	\quad &+v^{n + 1}_{q, j-1} \left(-\frac{\Delta \tau}{2\Delta s^2} \sigma^2 + (\mu- s)\frac{\Delta \tau}{\Delta s}1_{\mu <  s}\right)
	= v^n_{q, j} - \frac{\sigma^2}{2}\frac{\Delta \tau}{4\Delta s^2}(v^{n+1}_{q, j + 1} - v^{n+1}_{q, j - 1})^2\\
	\quad &+\frac{A\Delta\tau}{\kappa + 1} \left(1+\frac{1}{\kappa}\right)^{-\kappa}\left[e^{-\kappa(-s - v^{n+1}_{q-1, j} + v^{n+1}_{q, j})}  
	+  e^{-\kappa(s - v^{n+1}_{q+1, j} + v^{n+1}_{q, j})}\right].
\end{split}
\end{equation}
where $s$ is the value of the $j$-th grid point in the discretized reference-price space.

We iteratively solve this non-linear equation for $v^{n+1}_{q, j}$. Let $v^{n+1, k}_{q, j}$ denote the solution of this system in $k$-th iteration. Then in 
$(k+1)$-th iteration, we construct a linear equation for $v^{n+1, k+1}_{q, j}$ by replacing all the $v^{n+1}_{q, j}$ on the right hand side of 
\eqref{finite difference discretized PDE} with $v^{n+1, k}_{q, j}$ and all  $v^{n+1}_{q, j}$ on the left 
hand side of \eqref{finite difference discretized PDE} with $v^{n+1, k+1}_{q, j}$. We repeat this procedure until the difference between $v^{n+1, k+1}_{q, j}$ and 
$v^{n+1, k}_{q, j}$ becomes negligible for all $q$ and $j$. Once the iteration converges, we can go on and compute $v^{n+2}_{q, j}$, otherwise our scheme breaks down.

\subsubsection{Boundary condition}\label{sec: Boundary Condition}

In numerical experiments, we used two different boundary conditions:
\paragraph{Assumption of the total amount of asset}
We assume that the quantity of the asset $q$ must be between $-Q$ and $Q$: when $q=Q$, buying is forbidden and selling is forbidden when $q=-Q$.

\paragraph{Zero second-derivatives at boundaries}
We also use a different boundary condition for the finite difference method, which is exact at the terminal time:
\begin{equation}
	v^{n}_{Q, j} - v^{n}_{Q-1, j}  = v^{n}_{Q-1, j} - v^{n}_{Q-2, j},\quad v^{n}_{-Q, j} - v^{n}_{-Q+1, j}  = v^{n}_{-Q+1, j} - v^{n}_{-Q+2, j}.
\end{equation}
\newline
From our experiments, we observe that
\begin{itemize}
\item Different boundary conditions will eventually have an impact on the solution even at points away from the boundaries when $T$ is sufficiently large.
\item The ``Zero second-derivative'' boundary condition  ``pushes'' the optimal limit prices to constants, while the assumption of the finite total amount of asset would ``prevent'' the optimal limit prices from converging to constants.
\item To observe the ``true'' phenomena when time is far away from terminal, we need to set the space of inventory large enough. In our experiments, when we set the numerical boundaries for inventory to be $\pm1000$, the results corresponding to different boundary conditions match very well for $q$ between $\pm 300$, and they both indicate the same convergence of the optimal limit prices when time is away from terminal.
\end{itemize}

\subsection{Split-step method}
We introduce the split-step method widely used to solve nonlinear Schr\"odinger equations. We make a change of variables 
$u(t,q,x,s) = -e^{-x}\tilde{v}(t,s,q)$ for $u$ in \eqref{very first value function}.  Then for $-Q<q<Q$,
\begin{equation*}\label{equation tilde v first}
	\tilde{v}_t(t,s,q) + (\mu-s)\tilde{v}_s(t,s,q) + \frac{1}{2}\sigma^2 \tilde{v}_{ss}(t,s,q) 
	- \frac{A}{\kappa+1}\left(e^{-\kappa\delta^b}+e^{-\kappa\delta^a}\right)\tilde{v}(t,s,q) = 0
\end{equation*}
and
\begin{align*}
	& \tilde{v}_t(t,s,-Q) + (\mu-s)\tilde{v}_s(t,s,-Q) + \frac{1}{2}\sigma^2 \tilde{v}_{ss}(t,s,-Q) 
	- \frac{A}{\kappa+1}e^{-\kappa\delta^b}\tilde{v}(t,s,-Q) = 0\\
	& \tilde{v}_t(t,s,Q) + (\mu-s)\tilde{v}_s(t,s,Q) + \frac{1}{2}\sigma^2 \tilde{v}_{ss}(t,s,Q) 
	- \frac{A}{\kappa+1}e^{-\kappa\delta^a}\tilde{v}(t,s,Q) = 0 
\end{align*}
with the optimal feedback control
\begin{align*}
	\delta^b(t,s,q) &=  s + \left[\log\left(1+\frac{1}{k}\right)+\log \tilde{v}(t,s,q+1)-\log \tilde{v}(t,s,q)\right]\\
	\delta^a(t,s,q) &= -s + \left[\log\left(1+\frac{1}{k}\right)+\log \tilde{v}(t,s,q-1)-\log \tilde{v}(t,s,q)\right],
\end{align*}
and the terminal condition $\tilde{v}(T,s,q) = e^{- sq}$.

Let the vector
\begin{equation}
	\vec{v}(t,s) = \left[\tilde{v}(t,s,-Q), \tilde{v}(t,s,-Q+1), \ldots, \tilde{v}(t,s,Q-1), \tilde{v}(t,s,Q)\right]^T.
\end{equation}
Then the evolution equation for $\tilde{v}(t,s,q)$ can be written as
\begin{equation}
	\vec{v}_t(t,s) + \mathcal{S}\vec{v}(t,s) + \mathcal{Q}\vec{v}(t,s) = 0,
\end{equation}
where $\mathcal{S}$ and $\mathcal{Q}$ are operators applied to $\vec{v}$:
\begin{equation}
	\mathcal{S}\vec{v}(t,s) = \left[(\mu-s)\frac{\partial}{\partial s} + \frac{1}{2}\sigma^2 \frac{\partial^2}{\partial s^2}\right]\vec{v}(t,s)
\end{equation}
\begin{equation}
	\mathcal{Q}\vec{v}(t,s) = - \frac{A}{\kappa+1}
	\begin{pmatrix}
		e^{-\kappa\delta^b} \tilde{v}(t,s,-Q)\\
		(e^{-\kappa\delta^a}+e^{-\kappa\delta^b})\tilde{v}(t,s,-Q+1)\\
		\vdots\\
		(e^{-\kappa\delta^a}+e^{-\kappa\delta^b})\tilde{v}(t,s,Q-1)\\
		e^{-\kappa\delta^a} \tilde{v}(t,s,Q)
	\end{pmatrix}
\end{equation}
The idea of the split-step method is to solve the $\mathcal{S}$ and $\mathcal{Q}$ evolutions separately:
\begin{equation}
	\label{eq:separate evolution equations}
	\vec{v}_t(t,s) + \mathcal{Q}\vec{v}(t,s) = 0,\quad 
	\vec{v}_t(t,s) + \mathcal{S}\vec{v}(t,s) = 0.
\end{equation}
The formal solutions to the separate evolution equations are $\vec{v}(t,s) = e^{(T-t)\mathcal{Q}}\vec{v}(T,s)$ and 
$\vec{v}(t,s) = e^{(T-t)\mathcal{S}}\vec{v}(T,s)$, respectively.  For each time step, we first consider the $\mathcal{Q}$ evolution and then consider 
the $\mathcal{S}$ evolution.  The formal solution of such a scheme is then given by
\begin{equation}
	\vec{v}(t_{n+\frac{1}{2}},s) = e^{-\Delta t\mathcal{Q}}\vec{v}(t_{n+1},s),\quad
	\vec{v}(t_n,s) = e^{-\Delta t\mathcal{S}}\vec{v}(t_{n+\frac{1}{2}},s),
\end{equation}
or equivalently,
\begin{equation}
	\vec{v}(t_n,s) = e^{-\Delta t\mathcal{S}}e^{-\Delta t\mathcal{Q}}\vec{v}(t_{n+1},s),
\end{equation}
and we expect that $\vec{v}(t_n,s)$ converges to the real solution as $\Delta t\to 0$.

The advantage of the split-step method is that we are able to solve the separate evolution equations in (\ref{eq:separate evolution equations}) easily and therefore 
we can numerically solve the full evolution equation effectively.

Moreover, to avoid numerical overflow or underflow issue, for each time step $t_n$, we can utilize the scale-invariance of the differential equation and normalize 
the numerical solution by taking $\tilde{v}(t_n, s_i, q)/c_n$ with an appropriate constant $c_n>0$.

\subsubsection{The $\mathcal{Q}$ evolution}

We note that the equation $\vec{v}_t(t,s) + \mathcal{Q}\vec{v}(t,s) = 0$ is a special case of the Avellaneda and Stoikov model 
\cite{Avellaneda06high-frequencytrading}, and the equation is solvable by the technique described in \cite{InventoryRisk,zhang2009two}.  
Let $w(t,s,q) = e^{-\kappa s q} \tilde{v}^{-\kappa}(t,s,q)$, and then $w$ satisfies
\begin{align*}
	& w_t(t,s,q) + \frac{A\kappa}{\kappa+1}\left(1+\frac{1}{\kappa}\right)^{-\kappa}[w(t,s,q-1)+w(t,s,q+1)] = 0,\quad -Q<q<Q\\
	& w_t(t,s,-Q) + \frac{A\kappa}{\kappa+1}\left(1+\frac{1}{\kappa}\right)^{-\kappa}w(t,s,-Q+1) = 0\\
	& w_t(t,s,Q) + \frac{A\kappa}{\kappa+1}\left(1+\frac{1}{\kappa}\right)^{-\kappa}w(t,s,Q-1) = 0.
\end{align*}
The above equations are solvable and for a fixed $s$,
\begin{equation*}
	\vec{w}(t,s) = e^{\Delta t M}\vec{w}(t+\Delta t,s),
\end{equation*}
where $\vec{w}(t,s) = (w(t,s,-Q), w(t,s,-Q+1), \ldots, w(t,s,Q-1), w(t,s,Q))^T$ and $M$ is a tridiagonal matrix with diagonal elements equal to 0 and off-diagonal elements equal to $\frac{A\kappa}{\kappa+1}\left(1+\frac{1}{\kappa}\right)^{-\kappa}$. We let $\tilde{v}(t,s,q) = e^{-s q} w^{-\frac{1}{\kappa}}(t,s,q)$ after
$w(t,s,q)$ is obtained numerically.

%

\subsubsection{The $\mathcal{S}$ evolution}

We now consider the $\mathcal{S}$ evolution: $\vec{v}_t(t,s) + \mathcal{S}\vec{v}(t,s) = 0$.  Note that $\mathcal{S}$ is an OU operator. Therefore, by the Feynman-Kac formula, 
\begin{equation}
	\tilde{v}(t,s,q) = \mathbb{E}[\tilde{v}(t+\Delta t, S_{t+\Delta t}, q)|S_t=s],\quad
	dS_t = (\mu - S_t)dt + \sigma dB_t,
\end{equation}
where $B_t$ is a Brownian motion.

To compute the numerical expectation, We discretize the $s$ domain uniformly: $S_{\min}=s_1, s_2, \ldots, s_{N-1}, s_N = S_{\max}$ and $s_{i+1}-s_i=\Delta s$,
and let 
\begin{align*}
	& \tilde{v}(t,s_i,q) = \sum_{j=1}^{N}p_{ij}\tilde{v}(t+\Delta t,s_{j},q)\\
	& p_{ij} = \mathbb{P}(s_j-\Delta S/2 < S_{t+\Delta t} \leq s_j + \Delta S/2|S_t=s),\quad 2\leq j\leq N-1\label{MC transition probability}\\
	& p_{i1} = \mathbb{P}(S_{t+\Delta t} \leq s_1 + \Delta S/2|S_t=s),\quad p_{iN} = \mathbb{P}(s_N-\Delta S/2 < S_{t+\Delta t}|S_t=s).
\end{align*} 
Given that $S_t=s$, $S_{t+\Delta t}$ is a Gaussian random variable with mean $e^{-\Delta t}s+(1-e^{-\Delta t})\mu$
and variance $\frac{\sigma^{2}}{2}(1-e^{-2\Delta t})$, we can easily compute the numerical values of $p_{ij}$.

Here we approximate the OU process $S_t$ by a discrete time, discrete space Markov chain.  To ensure the accuracy of the numerical expectation,
we compare the stationary distributions of $S_t$ and the Markov chain, and we calibrate $\Delta s$ and $\Delta t$ so that those two stationary
distributions are close.  To do that, we first select a reasonable step-size $\Delta t$ for the time space (in our experiments, we typically choose $1/10$ or 
$1/100$ of the mean reversion time). Then we choose a sufficiently small $\Delta s$ such that the difference between the two stationary distributions are negligible.

\bibliographystyle{plain}
\bibliography{reference}

\end{document}